\documentclass[prb,twocolumn,showpacs,superscriptaddress]{revtex4}
\usepackage{graphicx,amsmath,amssymb}

\begin{document}

\title{Depletion induced isotropic-isotropic phase separation in suspensions of rod-like colloids}

\author{S. Jungblut}
\affiliation{Institut f\"ur Physik, Johannes Gutenberg-Universit\"at,
D-55099 Mainz, Staudinger Weg 7, Germany}

\author{R. Tuinier}
\affiliation{Forschungszentrum J\"ulich, Institut f\"ur
Festk\"orperforschung, D-52425 J\"ulich, Germany}

\author{K. Binder}
\affiliation{Institut f\"ur Physik, Johannes Gutenberg-Universit\"at,
D-55099 Mainz, Staudinger Weg 7, Germany}

\author{T. Schilling}
\affiliation{Institut f\"ur Physik, Johannes Gutenberg-Universit\"at,
D-55099 Mainz, Staudinger Weg 7, Germany}

\date{\today}

\begin{abstract} 
When non-adsorbing polymers are added to an isotropic suspension of rod-like
colloids, the colloids effectively attract each other via depletion forces. 
We performed Monte Carlo simulations to study the phase diagram of 
such rod-polymer mixture. The colloidal rods were modelled as hard
spherocylinders; the polymers were described as spheres of the same diameter
as the rods. The polymers may overlap with no energy cost, while overlap of
polymers and rods is forbidden.   

Large amounts of depletant cause phase separation of the mixture. We estimated
the phase boundaries of isotropic-isotropic coexistence both, in the 
bulk and in confinement. To determine the
phase boundaries we applied the grand canonical ensemble using successive 
umbrella sampling [J. Chem. Phys. {\bf 120}, 10925 (2004)], and we performed a
finite-size scaling analysis to estimate the location of the critical point. 
The results are compared with predictions of the free volume
theory developed by Lekkerkerker and Stroobants [Nuovo Cimento D {\bf 16}, 949
(1994)]. We also give estimates for the interfacial tension between the
coexisting isotropic phases and analyse its power-law behaviour on approach 
of the critical point. 
\end{abstract}

\pacs{82.70.Dd, 64.70.Ja, 68.35.Rh, 05.10.Ln}
\maketitle
\section{Introduction}
Non-adsorbing polymers are often added to colloidal suspensions in order to
modify the effective interactions between the colloids. By this means
phase transitions can be driven, e.\ g., in order to facilitate the 
isolation of the colloids \cite{hebert:1963}.
In particular, mixtures of viruses and polymers are widely used in experiments
on colloidal liquid crystals (see, for instance, the recent review by Dogic
and Fraden \cite{dogic:2006}). 

As certain viruses, such as Tobacco Mosaic Virus and fd-virus, can to a first
approximation be regarded as long cylinders, they are good model systems for 
liquid crystals. Currently there is much interest in the
non-equilibrium behaviour of these systems -- in particular in the effect of
shear \cite{kang:2006, dhont:2003, lenstra:2001, lettinga:2004}. 
In order to interpret the
experimental results, detailed theoretical knowledge on the equilibrium 
phase diagram is needed. Therefore we present in this article a computer 
simulation study of the phase behaviour of rods and spheres, both in the 
bulk and in confinement.
    
The Asakura-Oosawa-Vrij model \cite{asakura.oosawa:1954, vrij:1976} is 
a useful model for mixtures of polymers and 
spherical colloids. The polymers are assumed to be freely interpenetrable 
with respect to each other, while there is a hard-core
repulsion between the colloids as well as between the colloids and the 
polymers. As the interaction energy is then either zero or infinite, the 
phase behaviour is purely of entropic origin. 

In this article we discuss a similar model for mixtures of rod-like colloids
and polymers: a mixture of hard spherocylinders with length $L$ and diameter
$D$ and freely interpenetrable spheres with the same diameter.   

This system was studied with liquid-state integral equation theory 
  \cite{chen:2002, chen:2004, cuetos:2007} and free volume
  theory \cite{lekkerkerker:1994} as well as with computer simulations
  \cite{li:2005, bolhuis:1997, savenko:2006}.
In recent years various regimes of size-ratios and concentrations
have been discussed, e.\ g.\, the packing properties at very high 
concentrations or the behaviour of small rods, which act as a depletant on 
large spheres. 
Here we focus on dispersions of rods and spheres of similar diameters at low
  concentrations.
The phase diagram of this model was studied within free volume theory by
Lekker\-ker\-ker and Stroobants \cite{lekkerkerker:1994}. 
Details of this approach will be described in section \ref{theory}.
Li and Ma recently computed the effective interaction between two rods by
Monte Carlo simulation \cite{li:2005}.
Bolhuis and coworkers\cite{bolhuis:1997} as well as Savenko and Dijkstra
\cite{savenko:2006} determined the bulk rod-sphere phase diagram by  
simulation (via thermodynamic integration). Both have given results 
for rods of aspect ratio
$L/D=5$ and spheres of several 
diameter values. In order to avoid simulating the spheres explicitly, 
Savenko and Dijkstra used an exact
expression for the effective Hamiltonian, which was numerically
evaluated during the Monte Carlo simulation for each rod configuration.  
Bolhuis and coworkers\cite{bolhuis:1997} modelled the spheres
explicitly in their Gibbs Ensemble Monte Carlo simulations to study
fluid-fluid separation, while for the other parts of the phase diagram 
they used an effective expression for the rod-rod interaction.
 
Here we present results for the full rod-sphere system, which were 
obtained in the grand canonical ensemble.
The successive umbrella sampling method \cite{virnau:2004} was employed to 
determine the grand potential hypersurface of the system. This allowed us to
predict the phase boundaries of isotropic-isotropic coexistence with much
higher accuracy than the studies mentioned above, which used thermodynamic 
integration. In particular, we determined the 
critical point by a finite-size scaling analysis. We also studied the
influence of confinement on the phase behaviour. This is relevant
for experiments under shear.

In section \ref{theory} we briefly review free volume theory for the
rod-sphere mixture. In section \ref{sims} the simulation method is introduced. 
In section \ref{sec:res} we show results for phase diagrams and interfacial
tensions and compare them to the theoretical predictions. 
Section \ref{summary} contains a summary and a discussion.   

\section{Free volume theory}
\label{theory}
We briefly review the theoretical approach to rod-sphere
mixtures which was introduced by Lekkerkerker and Stroobants
\cite{lekkerkerker:1994} in 1994. 
The starting point is the thermodynamic potential in the
semi-grand canonical ensemble, where the number of rods is fixed, while the
 number of ``penetrable hard'' spheres is set by the chemical potential of
 spheres in a hypothetical reservoir that is held in equilibrium with the
 system. The potential $\Omega^{\rm
   semi}(N_r,V,T,{\mu}_s)$ of such a mixture can be written as: 
\begin{eqnarray}
\Omega^{\rm semi}&=&F(N_r,V,T)
  +\int_{-\infty}^{\mu^R_s}\frac{\partial\Omega(N_r,V,T,\mu^{'R}_s)}{\partial
  \mu^{'R}_s}d\mu^{'R}_s\nonumber \\
&=& F(N_r,V,T)-\int_{-\infty}^{\mu^R_s}N_s(\mu^{'R}_s) d\mu^{'R}_s,
\end{eqnarray}
where $F(N_r,V,T)$ is the free energy of $N_r$ rods in a volume $V$ at
temperature $T$, and the second term
accounts for the perturbation due to the addition of $N_s$ spheres 
from a reservoir, where the chemical potential is kept at $\mu^R_s$.
Osmotic equilibrium requires the chemical
potentials of the system and the reservoir to be equal. The chemical potential
of an ideal gas of spheres is connected to the density $\rho_s$ via 
\begin{equation}
 \rho_s=\exp \left( \frac{\mu_s}{k_BT} \right).
\end{equation} 
 Thus, the
number of spheres in the system depends on $\mu^R_s$. The only influence
the rods have on the spheres is the reduction of the free volume:
\begin{equation}
 N_s=\rho ^R_sV_{\rm free},
\end{equation} 
where $V_{\rm free}$ is the volume accessible for spheres under the
assumption that the rod configurations are undistorted upon adding spheres and
$\rho ^R_s$ is
the number density of spheres in the reservoir. The free
volume fraction $\alpha$ is defined as: 
\begin{equation}
\alpha = \frac{V_{\rm free}}{V}
\end{equation} 
and it can be calculated within scaled particle theory\cite{lekkerkerker:1994}.
Hence, the expression for the semi-grand canonical potential is reduced to: 
\begin{equation}
 \Omega^{\rm semi} = F(N_r,V,T) - \alpha V k_BT  \rho^R_s.
\end{equation}  
The chemical potential as well as the osmotic pressure $\Pi_r$ 
of the rods in the mixture can be
obtained from the thermodynamic relationships:
\begin{eqnarray}
\mu_r&=&\left ( \frac{\partial \Omega^{\rm semi}}{\partial N_r}
\right )_V \label{chem}\\
\Pi_r& = &-\left ( \frac{\partial
    \Omega^{\rm semi}}{\partial V} \right )_{N_r} \label{pressure} 
\end{eqnarray} 
where $v_r = \pi D^3(2+3L/D)/12$  is the volume of a spherocylinder of 
length $L$ and diameter $D$.

The virial expansion of the free energy of a system of hard spherocylinders 
can be calculated using scaled particle theory\cite {cotter:1977}:  
\begin{equation}\label{freeenergy}
\frac{Fv_r}{k_BTV} = \frac{y}{y+1} \left( {\rm const} + \ln(y) + \sigma [f] + \Pi_2y +
\frac{\Pi_3}{2} y^2 \right)
\end{equation}
with 
\begin{equation}
y = \frac{\eta_r}{1-\eta_r} \quad , 
\end{equation}
where $\eta_r = \rho_r v_r$ is the volume fraction of rods.

\begin{eqnarray}\label{pcoeffs}
\Pi_2& =& 3+\frac{3(\tau-1)^2}{3\tau-1}\rho[f]\nonumber \\
\Pi_3& =& \frac{12\tau(2\tau-1)}{(3\tau-1)^2} +
\frac{12\tau(\tau-1)^2}{(3\tau-1)^2}\rho[f] \quad ,
\end{eqnarray}
and $\tau$ is the overall length-to-diameter ratio of the spherocylinders
\begin{equation}
\tau = \frac{L+D}{D} \quad .
\end{equation}
In eqs. \ref{freeenergy} and \ref{pcoeffs}, $\sigma [f]$ and $\rho[f]$ are
functionals of the orientational distribution
function $f$: 
\begin{eqnarray}
\sigma[f]& =& \int f( {\bf u } )\ln[4\pi f( {\bf u})] d{\bf u} \label{functs} \\
\rho[f]& = &\frac {4}{\pi} \int \int \sin[\Phi({\bf u}, \label{functr} {\bf
u'})] f({\bf u})f({\bf u'}) d {\bf u} d {\bf u'}
\end{eqnarray}
where {\bf u} is the unit vector along a particle's axis.

The normalised orientational distribution function of a single rod $f$ is
adapted to minimise the
semi-grand canonical potential. In the nematic case it can be obtained either
numerically \cite{vanroij:2005} or from the Gaussian approximation
\cite{lekkerkerker:1994}, but here we are interested in the isotropic case,
i.\ e., all orientations ${\bf u}$ are equally probable and the distribution is
uniform:   

\begin{equation}
f({\bf u}) = \frac{1}{4\pi} \quad .
\end{equation} 
Therefore, in our case, the functionals are reduced to:
\begin{eqnarray}
 \sigma[f]& = &0 \\
 \rho[f]& = &1 
\end{eqnarray}
The expression for the free volume fraction calculated with scaled
particle theory reads:  
\begin{equation}\label{freevolume}
\alpha = \frac{1}{1+y} e^{-\left( Ay+By^2+Cy^3 \right)},
\end{equation}
where, in the isotropic case, for ``penetrable hard'' spheres with diameter
$D$: 
\begin{eqnarray}
A&=&\frac{6\tau}{3\tau-1}+\frac{3(\tau+1)}{3\tau-1}+\frac{2}{3\tau-1}\nonumber
\\
B&=&\frac{18\tau^2}{(3\tau-1)^2}+\left
  (\frac{6}{3\tau-1}+\frac{6(\tau-1)^2}{(3\tau-1)^2}\right ) \nonumber \\
C&=&\frac{2}{3\tau-1}\left (
  \frac{12\tau(2\tau-1)}{(3\tau-1)^2}+\frac{12\tau(\tau-1)^2}{(3\tau-1)^2}\right ) \label{abc}
\end{eqnarray}

Thus, the chemical potential and the osmotic pressure of the rods in the
mixture are
functions of the rod volume fraction and of the chemical potential of spheres
in the reservoir. At coexistence, they are equal. From this condition we have 
obtained the phase diagrams, which are compared to our simulation results in
section \ref{sec:res}. 
 
\section{Model and simulation method}
\label{sims} 
In our Monte Carlo simulations we model the colloids as hard 
spherocylinders of length $L$ and diameter $D$. 
The polymers are approximated as spheres of the
same diameter, which are hard with respect to the
rods and freely interpenetrable among each other. 

\begin{figure}
\begin{center}
\includegraphics[clip=,width=0.9\columnwidth]{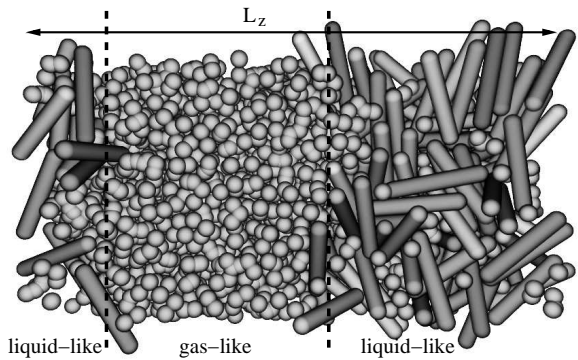} 

\caption{\label{snapshot} Configuration snapshot for $L/D=5$ at coexistence.} 
\end{center}
\end{figure}

We performed simulations in the grand canonical ensemble, where the volume 
$V$ and the chemical potentials $\mu_r$ of the spherocylinders and
$\mu_s$ of the spheres are fixed. The temperature $T$ is formally also fixed,
but irrelevant, since it only sets the energy scale. 
Simulations were performed in a box with edges $L_z \ge L_y =
L_x \ge 3L$ and periodic boundary conditions. In a rectangular box
configurations at coexistence form preferably such that the interfaces are
parallel to the small faces of the box (see figure \ref{snapshot}). 
This simplifies the analysis of the
interfacial tension. The finite-size effects were examined in a cubic box.  
For the study of confinement effects on our model system we choose a geometry
where the box dimensions were $L_x = L_y \ge 3L$ and $L_z = 3L$ with
periodic boundary conditions in the $x$- and $y$-directions and hard
  walls in the $z$-direction.
To speed up the simulations we employed a cell system for efficient overlap 
detection of anisotropic particles\cite{vink:2005}. In this approach 
the simulation box is cut into cubes of side lengths $\ge D$. Whether a rod 
intersects a cube or not can be computed very fast. The volume which needs 
to be checked for overlaps then contains at the most a few particles.

\subsection{Cluster move}

In order to efficiently equilibrate canonical simulations of spherical 
colloids, Biben and co-workers introduced a cluster move\cite{biben:1996}. 
Later this move was extended to simulations in the grand canonical 
ensemble \cite{vink:2004}. Here we introduce a version that is adjusted
to the case of spherocylinders. 

If we attempt to insert a spherocylinder into a simulation box 
full of spheres,
it will certainly overlap with some of them and the move will be
rejected. The cluster move combines the insertion of a spherocylinder with 
the removal of the spheres, which it otherwise would overlap with. 
Thus it increases the probability of accepting an insertion move.
At the same time, the removal of a
spherocylinder is combined with the insertion of a certain amount of spheres
in the void left by the removed spherocylinder. 
 
Care has to be taken of the acceptance probabilities in order to ensure
detailed balance. Here, we only give the most relevant equations. 
The reader who is interested in details is referred to the work by 
Vink and Horbach \cite{vink:2004}, who have explained 
the process thoroughly for spherical colloids. For spherocylinders we have to
account for an additional
degree of freedom -- the orientation.     
The Metropolis probability of accepting a removal of a spherocylinder in
a cluster move is:
\begin{equation}
\label{eq:remove}
acc(N_r \rightarrow N_r-1) = \min \left [1, \frac{N_r m
    (z_s V_{\rm dz})^{n_s}}{4 \pi V z_r n_s!}e^{-\beta \Delta E} \right ]
\end{equation}
where $\Delta E$ is the potential-energy difference between the initial and
the final configuration -- which is infinite for an overlapping
configuration and zero otherwise. The volume of the depletion zone around
a spherocylinder is $V_{\rm dz}=\pi(2D)^3(3L/(2D)+2)/12$. 
The factors $1/(4\pi)$ in eq.\ \ref{eq:remove} and $4\pi$ in
eq.\ \ref{eq:insert}
are due to the orientational degree of freedom. The fugacities $z_{r/s}$ of
spherocylinders and
spheres respectively are related to the chemical potentials via $z_{r/s} =
\exp(\beta \mu _{r/s})$, where $\beta =1/(k_BT)$ is the inverse temperature. 
$n_s$
is the number of spheres to be inserted into the void. $n_s$ is drawn uniformly
from the interval $[0,m \rangle $, where $m$ is an integer given by
$m=1+\max[1,$ int$(z_sV_{\rm dz}+a \sqrt{z_sV_{\rm dz}})]$, with $a$
a positive constant of order unity.  

If only one spherocylinder is removed and no spheres are inserted, 
the acceptance probability is reduced to: 
\begin{equation}
acc(N_r \rightarrow N_r-1) = \min \left [1, \frac{N_r}{4
    \pi V z_r}e^{-\beta \Delta E} \right ]
\end{equation}

The Metropolis probability of accepting an insertion of a spherocylinder in a
cluster move is:
\begin{equation}
\label{eq:insert}
acc(N_r \rightarrow N_r+1) = \min \left [1,\frac{z_r
    n_s! 4 \pi V e^{-\beta \Delta E} }{(V_{\rm dz}z_s)^{n_s} m (N_r+1)} \right ],
\end{equation}
if $n_s < m$ and $0$ otherwise. Here $n_s$ is the number of spheres which need 
to be removed to produce a void, in which the spherocylinder 
can be inserted.    
If only one spherocylinder is inserted and no spheres are removed, 
the acceptance probability is reduced to: 
\begin{equation}
acc(N_r \rightarrow N_r+1) = \min \left [1, \frac{z_r 4
    \pi V}{N_r+1}e^{-\beta \Delta E} \right ]
\end{equation}

\subsection{Phase boundaries from grand-canonical simulations}
We determined the phase boundaries from the probability
distribution $P(N_r)$, which is the probability to
observe $N_r$ rods in the mixture for given chemical potentials $\mu_r$ and
$\mu_s$. For a given value of $\mu_s$, we searched for the value of 
$\mu_r$ at which the distribution is bimodal.
However, the simulations need not
be performed right at the coexistence chemical potential (which is usually not
known beforehand). In fact, the
chemical potential can be set to any value $\mu_r^{\rm sim}$ and
then renormalised to coexistence via 
\begin{equation}
\ln [P_{\mu_r}(N_r)]=\ln [P_{\mu_r^{\rm sim}}(N_r)] + (\mu_r - \mu_r^{\rm
  sim})N_r
\end{equation}
such that the areas under the two peaks in $P(N_r)$ are
equal\cite{landau.binder:2000}.
We used the successive umbrella sampling method
\cite{virnau:2004} to determine $P(N_r)$. This technique allowed us to 
sample regions between the two bulk phases where $P(N_r)$ is very low.

\begin{figure}[t]
\begin{center}
\includegraphics[clip=,width=0.9\columnwidth]{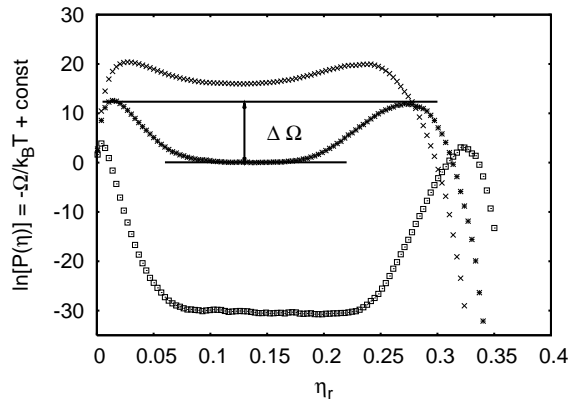} 

\caption{\label{5freeenergy} Grand potential 
  ${\rm ln}[P(\eta_r)]$ for spherocylinders of aspect ratio $L/D=5$ at sphere
  fugacities $z_s = 1.0125$ (crosses), $1.0625$ (stars) and $1.175$
  (squares). $\Delta \Omega$ indicates the height of the grand potential
  barrier between the coexisting phases.} 
\end{center}
\end{figure}

Figure \ref{5freeenergy} shows the logarithm of the probability distribution
$P(\eta_r)$ as a function of the rod volume fraction $\eta_r = N_rv_r/V$ with
$v_r = \pi D^3(2+3L/D)/12$, which -- up to an additive constant -- is equal to
the grand 
potential of the system. The locations of the maxima of the peaks are the bulk 
volume fractions of rods at coexistence. For a large fugacity $z_s$ (squares), 
and hence a large concentration of spheres, there are two clearly 
separated peaks indicating a phase 
transition which is strongly of first order. With decreasing
concentration of spheres (stars and crosses) the effective 
attraction between the 
rods becomes weaker. Hence the peaks broaden and move closer until eventually
the critical point is reached.

Figure \ref{snapshot} shows a typical configuration at coexistence. Due to the
periodic boundaries two slabs separated by two interfaces have formed.

In figure \ref{5freeenergy} one can also
clearly see that an advanced biasing scheme such as successive umbrella
sampling is necessary to bridge the huge differences in probabilities between
the pure bulk states and the states in the two-phase coexistence region (which
show up via a horizontal part of the $\Omega$ vs. $\eta_r$ curve, since a
change of $\eta_r$ just amounts to a change of the sizes of the
gas-like and liquid-like domains, without changing the areas of the interfaces
between them).

\section{Results}
\label{sec:res}
\subsection{Bulk}
\subsubsection{Phase diagram}

The phase diagram is presented choosing the fugacity of the
spheres and the rod volume fraction as variables. The fugacity $z_s$ is
related to the sphere reservoir volume fraction via $\eta_s^R=z_sv_s$, where
$v_s=\pi D^3/6$ is the volume of a sphere. Explicit implementation of the
spheres allows an easy transformation into the frame of ($\eta_s, \eta_r$),
where $\eta_s$ is the sphere volume fraction in the system. In free volume
theory the actual concentration of spheres follows as $\eta_s = \alpha
\eta_s^R$, with $\alpha$ from eq. \ref{freevolume}. 

\begin{figure}[t]
\begin{center}
\includegraphics[clip=,width=0.9\columnwidth]{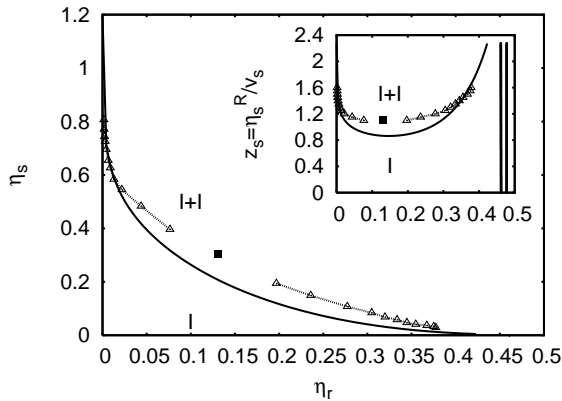} 

\caption{\label{bulk3} Bulk phase diagram for a mixture of spherocylinders with
  aspect ratio $L/D=3$ and spheres of diameter $D$. 
  The solid lines are predictions of the free volume
  theory. The filled square marks the critical point.
  The inset shows the phase diagram in the ($z_s, \eta_r$)-plane. The almost
  vertical lines in the inset indicate the coexistence region of the
  isotropic-nematic transition.} 
\end{center}
\end{figure}

\begin{figure}[t]
\begin{center}
\includegraphics[clip=,width=0.9\columnwidth]{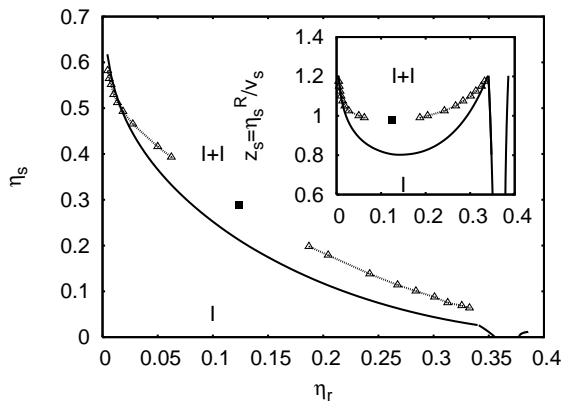} 

\caption{\label{bulk5} Bulk phase diagram for a mixture of spherocylinders with
  aspect ratio $L/D=5$ and spheres of diameter $D$. 
  The solid lines are predictions of the free volume
  theory. The filled square marks the critical point. The inset shows the
  phase diagram in the ($z_s, \eta_r$)-plane.The almost
  vertical lines in the inset indicate the coexistence region of the
  isotropic-nematic transition.} 
\end{center}
\end{figure}

Figures \ref{bulk3} and \ref{bulk5} show phase diagrams for mixtures of rods
with aspect ratios $L/D=3$ and $5$ and spheres. The solid lines are free
volume theory predictions. The single phase, the isotropic mixture of rods and 
spheres, is marked ``I''. The region of the phase diagram, where 
the gas-like and liquid-like isotropic phases coexist is named ``I+I''. The two
almost vertical lines are theoretical predictions for the phase boundaries
of the isotropic-nematic transition \cite{tuinier:2007}.

Since free volume theory is based upon a mean-field approximation, and
fluctuations, which are especially relevant near the critical point, are
ignored, we expect its predictions to deviate from the simulation results
there. Away from the critical point the predictions of the free volume theory 
approach the simulation results. As anti\-cipated, the theory underestimates the
volume fraction of spheres in the liquid phase
considerably. This is due to the fact that the depletion forces change the
local structure of the fluid \cite{schilling:2007} -- an effect which is not 
included in
the calculations of the free volume accessible to spheres. On the gas-branch
of the coexistence region, where the amount of rods is negligible, the
theoretical predictions agree well with the simulation results.

\begin{figure}[t]
\begin{center}
\includegraphics[clip=,width=0.9\columnwidth]{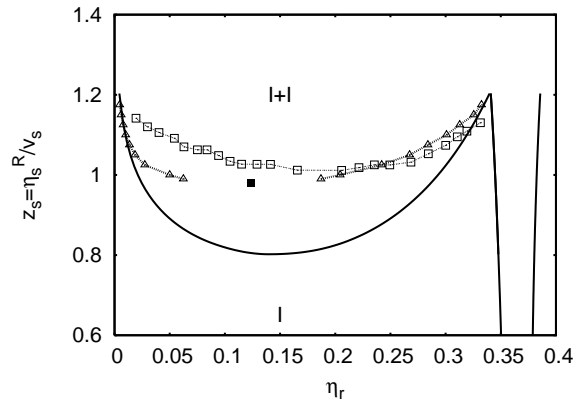} 

\caption{\label{bulk5comp} Comparison of the bulk phase diagram for a mixture
  of spherocylinders with
  aspect ratio $L/D=5$ and spheres of diameter $D$ to previous work. The
  open triangles are the
  results of the present work, the filled square marks the critical
  point. The open squares are the data obtained by Bolhuis and
  coworkers\cite{bolhuis:1997}. The solid lines are predictions of the free
  volume theory.} 
\end{center}
\end{figure}

Figure \ref{bulk5comp} shows a comparison of our results with the data
obtained in previous computer simulations on the fluid-fluid separation by
Bolhuis and coworkers\cite{bolhuis:1997} (open squares). The errorbars of
their data are $\Delta \eta_r \sim 0.02$. 
Thus, the results do not coincide within the
errorbars. We attribute the difference to the small system sizes
which were accessible at that time (1997). No estimate for the
critical point could be obtained from their work \cite{bolhuis:1997}.
The data from Savenko
and Dijkstra\cite{savenko:2006} does not lie within the range of this
graph. Presumably there is an error in reference [15]\footnote{Dr.\ S.\ 
Savenko, private communication to Dr.\ R.\ Vink.}. Therefore we cannot
compare to this work in detail.

The results presented here are of high accuracy: the errorbars are smaller
than the symbols.
The main sources of error are finite-size effects and insufficient 
sampling of the grand potential hypersurface perpendicular to the reaction 
coordinate (i.\ e., the packing fraction of rods). In order to check for the
quality of sampling, we repeated our simulations for several different
chemical potentials and different values of accuracy thresholds in the 
successive umbrella sampling. From this we estimate the error on the 
coexistence volume fractions to be $\Delta \eta_r = \pm 0.002$.
 
Finite-size effects enter in two ways: 1) At coexistence the system forms
slab configurations with two interfaces (due to the periodic boundaries). 
If $L_z$ is too small, the interfaces interact. Away from
the critical point, we ruled out this contribution by increasing $L_z$ 
such that a plateau appeared between the peaks in $P(\eta_r)$.
2) $P(\eta_r)$ depends on the lateral box size, because the 
spectrum of capillary waves on the interfaces is cut off for wavelengths larger
than the box. This effect is negligible
far away from the critical point, too, because the interfacial tension there
is large and the effects of capillary waves are very weak. However,
close to the 
critical point a finite-size scaling analysis becomes necessary. In fact, the
two effects are then just two aspects of the same property, the diverging 
correlation length. We discuss this issue further in 
subsection \ref{finitesize}.    

Close to the isotropic-nematic transition, the simulations cannot be
equilibrated properly with the methods used here. For accurate
grand-canonical simulations of the IN transition more advanced biasing
techniques are necessary\cite{vink.wolfsheimer:2005}. A study of this part of
the phase diagram is beyond the scope of the present work, however. 

\subsubsection{Finite-size scaling}
\label{finitesize}
\begin{figure}[t]
\begin{center}
\includegraphics[clip=,width=0.9\columnwidth]{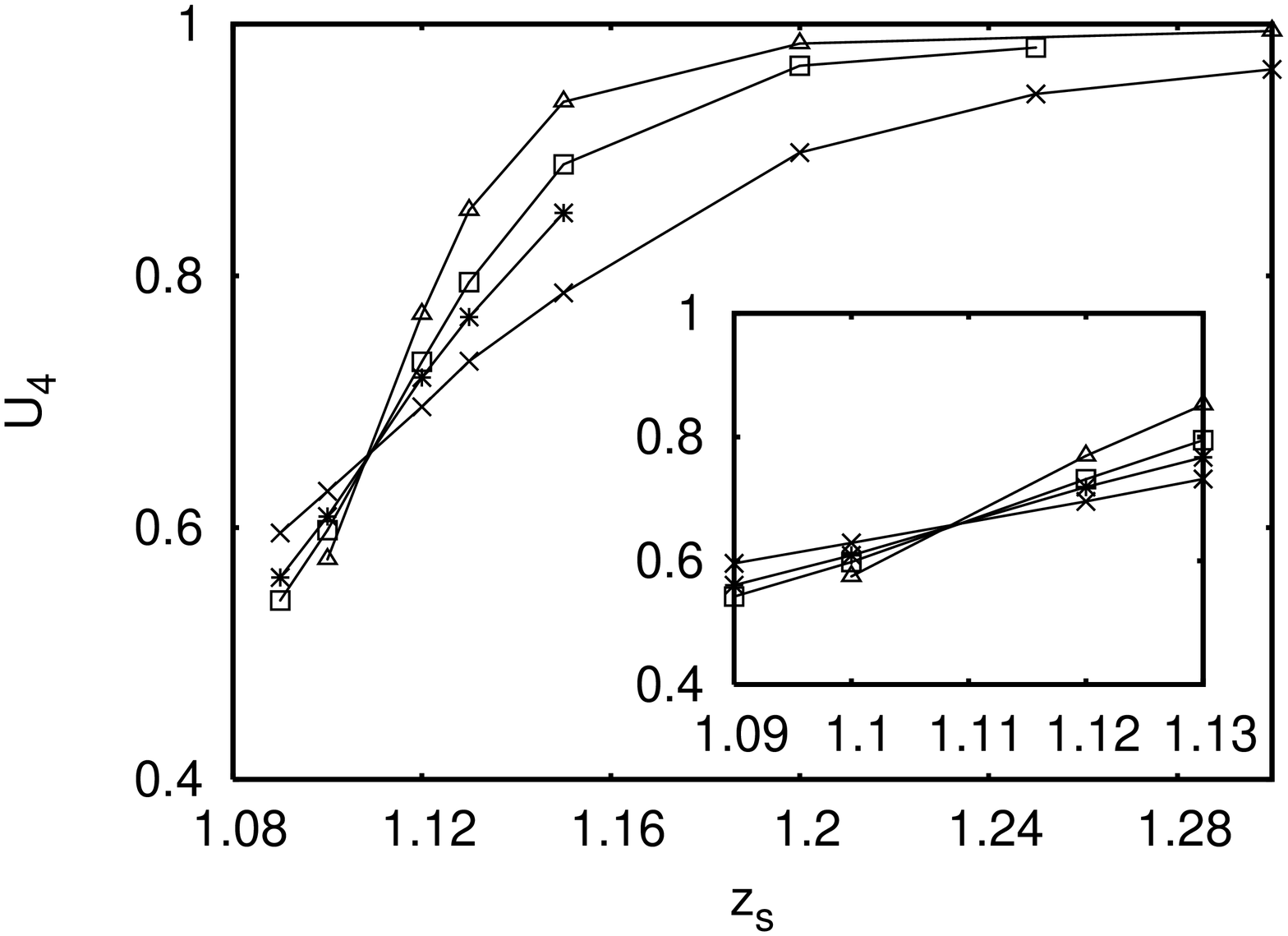} 

\caption{\label{u4ar3} Cumulant ratio $U_4$ as a function of the sphere
  fugacity close to the critical point for box lengths $L_x=9D$ (crosses), 
  $12D$ (stars), $14D$ (squares) and $18D$ (triangles).
  Spherocylinders aspect ratio is $L/D=3$. The inset shows a magnified plot of
  the region near the intersection point.} 
\end{center}
\end{figure}

As explained above, $P(\eta_r)$ depends on system size. This is, however, 
not a drawback but an advantage, because the finite-size effects can be used 
in order to locate the critical point of isotropic-isotropic
demixing. One can construct quantities which are independent of system 
size at the critical point. 
One possible choice \cite{landau.binder:2000, binder:1981} of
such a
quantity is the cumulant ratio
$U_4=\langle m^4 \rangle/\langle m^2 \rangle^2$ with
$m=\eta_r-\langle \eta_r \rangle$. 
When $U_4$ is plotted versus $z_s$ close to the critical point for different 
system sizes, the intersection point marks the critical point. 
Figure \ref{u4ar3} shows that the critical sphere fugacity for
spherocylinders with aspect ratio $L/D=3$ is $z_s^c = 1.109 \pm 0.001$.      

\begin{figure}
\begin{center}
\includegraphics[clip=,width=0.9\columnwidth]{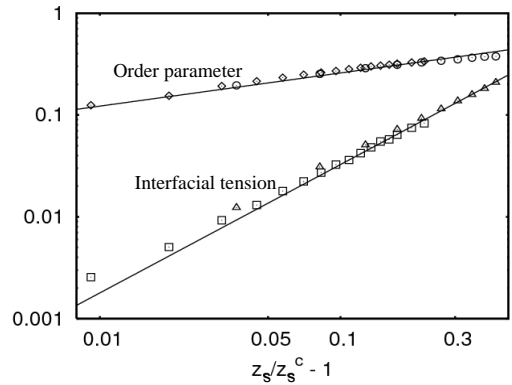} 

\caption{\label{critExp} Order parameter $(\eta_r^l - \eta_r^g)$ (circles and
  diamonds) and interfacial
  tension $\gamma$ (triangles and
  squares) versus
  ``relative distance from the critical point'' $z_s/z_s^c -
  1$. Spherocylinders aspect ratios are $L/D = 3$ (triangles and circles) and
  $5$ (squares and diamonds). The solid lines
  correspond to the Ising power-law-behaviour with exponents 
  $\beta_{\rm Ising} = 0.326$ and $2\nu_{\rm Ising} = 1.26$.} 
\end{center}
\end{figure}

Figure \ref{critExp} shows the difference in packing fraction $\eta_r^l -
\eta_r^g$ versus the ``relative distance from the critical point'' 
$z_s/z_s^c - 1$ (circles for $L/D = 3$ and diamonds for $L/D = 5$). The upper
line in the graph is proportional to $(z_s/z_s^c -
1)^{\beta_{\rm Ising}}$,
where $\beta_{\rm Ising} = 0.326$ is the critical exponent of the order
parameter in the Ising model. Clearly, the critical point is of the Ising
universality class. 
 
\subsubsection{Interfacial tension}

In the transition region where the gas phase transforms into the liquid phase
and vice-versa, a grand potential barrier $\Delta \Omega$ needs to be crossed 
(indicated in figure \ref{5freeenergy}). $\Delta \Omega$ is related to
the interfacial tension via
\begin{equation}
\gamma\equiv\lim_{L_x \rightarrow \infty} \frac{\Delta \Omega}{2L_x^2}
\end{equation}
where $L_x^2$ is the area of the interface and the factor $1/2$ accounts for
the two interfaces, which are present due to the periodic boundary
conditions\cite{binder:1982}.  
Figure \ref{critExp} shows values of the interfacial tension
as a function of the ``relative distance from the critical point'' 
$z_s/z_s^c - 1$ (triangles for $L/D = 3$ and squares for $L/D = 5$). The lower
line in the graph indicates 
$(z_s/z_s^c - 1)^{2\nu_{\rm Ising}}$, where $2 \nu_{\rm Ising} = 1.26$ 
is the critical exponent for the interfacial tension in the 3d Ising model.
 
\subsection{Confinement}

\subsubsection{Phase diagram}
\begin{figure}[t]
\begin{center}
\includegraphics[clip=,width=0.9\columnwidth]{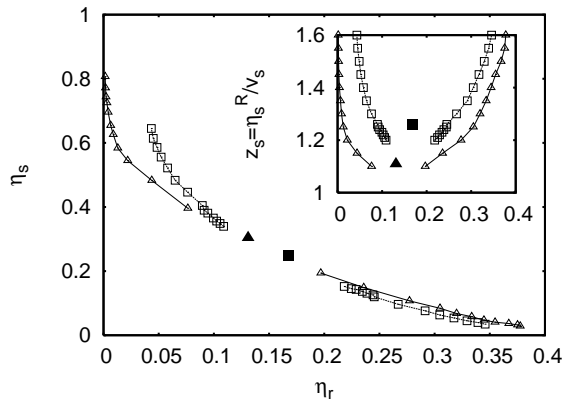} 

\caption{\label{conf3} Phase diagram for a mixture of spherocylinders with
  aspect ratio $L/D=3$ and spheres of diameter $D$ between two hard walls at
  distance $d/D = 3 L/D$ (squares) compared to the bulk values (triangles). 
  The filled symbols mark the critical points.
  The inset shows the phase diagram in the ($z_s, \eta_r$)-plane. The symbols
  show ``raw data'' for one box dimension $L_x = 12D$ only, and thus the
  curves marking the peaks of $P(\eta_r)$ do not join at the critical points
  (``finite size tails'' \cite{landau.binder:2000}).} 
\end{center}
\end{figure}

\begin{figure}[t]
\begin{center}
\includegraphics[clip=,width=0.9\columnwidth]{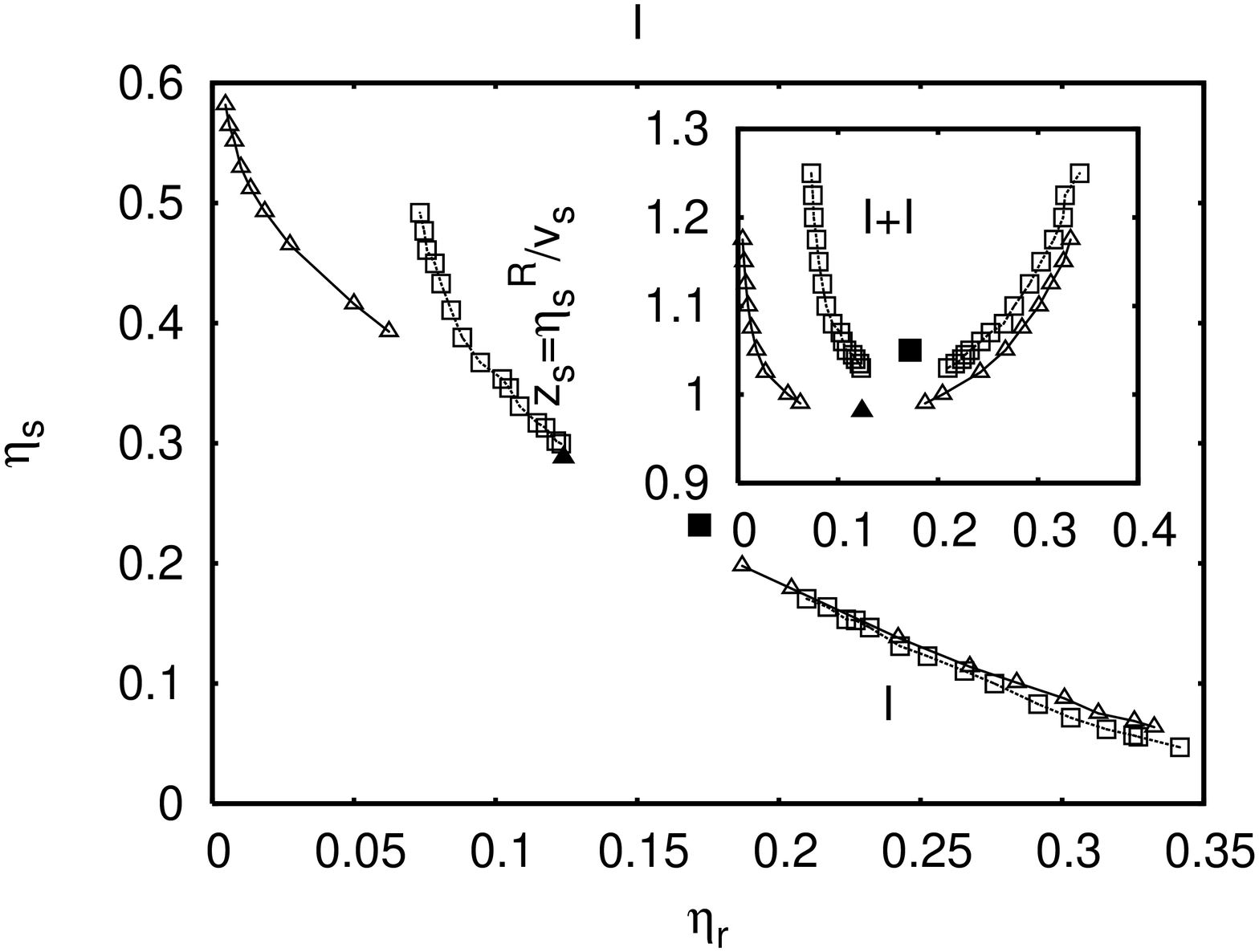} 

\caption{\label{conf5} Phase diagram for a mixture of spherocylinders with
  aspect ratio $L/D=5$ and spheres of diameter $D$ between two hard wall at
  distance $d/D = 3 L/D$ (squares) compared to the bulk values
  (triangles). The filled symbols mark the critical points.  
  The inset shows the phase diagram in the ($z_s, \eta_r$)-plane. The symbols
  show ``raw data'' for one box dimension $L_x = 20D$ only, and thus the
  curves marking the peaks of $P(\eta_r)$ do not join at the critical points
  (``finite size tails'' \cite{landau.binder:2000}).} 
\end{center}
\end{figure}

Now we consider the behaviour of the mixture confined between two hard
walls at distance $d$. Figures \ref{conf3} and \ref{conf5} show the
phase diagrams for rod aspect ratios $L/D =3$ and $5$. The distance
between the walls is $d/D = 3L/D$. Demixing in confinement
occurs at larger sphere fugacities than in the bulk. Also the chemical 
potential of the rods at coexistence is higher than in the bulk. 
The gas-like phase is shifted to larger
rod volume fractions. The amount of spheres in the system is smaller than in
the bulk at the same fugacity. The large increase of the concentration of rods
in the gas--like phase distinguishes this system clearly from the behaviour of
the Asakura-Oosawa-Vrij model in confinement. 
   
\begin{table}
\begin{tabular}{|c|c|c|c|c|}
\hline 
$L/D$ & $z_{s,{\rm conf}}^c$ & $\eta_{r, {\rm conf}}^c$ & $z_{s, {\rm bulk}}^c$ & $\eta_{r, {\rm bulk}}^c$ \tabularnewline
\hline
\hline 
$ 3 $ & $1.26 \pm 0.01$ & $0.168 \pm 0.002$ & $1.109 \pm 0.001$ 
& $ 0.131 \pm 0.002$\tabularnewline
\hline 
$ 5 $ & $1.05 \pm 0.01$ & $0.172 \pm 0.002$ & $0.981 \pm 0.001$
& $ 0.124 \pm 0.002$ \tabularnewline
\hline 
\end{tabular}
\caption{
  \label{table} 
  Critical point in the bulk and in confinement to slit-pore of width 
  $d/D=3L/D$.}
\end{table}

We also performed a finite size scaling analysis in confinement to see how the
critical point is shifted. Table \ref{table} lists the results in comparison to
the bulk results. 

The relative shift in $\eta_r^c$, which we observe for rods with aspect ratio
$L/D = 5$, is larger than the one of rods with aspect ratio $L/D = 3$, though
the relative shift in $z_s^c$ behaves the other way around. The second effect
is due to the ordering of the spheres close to the walls. Thus, its 
relative decrease on increase of the rods' aspect ratio is plausible, 
since, in terms of the sphere diameter, the distance between the walls 
increases.  

The shift in $\eta_r^c$ is caused by the wall-induced layering of the rods. 
Because of orientational ordering, this effect is much stronger than e.\ g.\ 
the shift in the Asakura-Oosawa-Vrij model, which is due to 
positional ordering \cite{vink:2006}. Close to the wall the rods are 
preferably oriented parallel to the
wall (``parallel anchoring''). Therefore, their concentration
is much higher than in the isotropic bulk.   

Obviously, it would be very interesting to
investigate how the critical point depends on the wall separation
$d$. This is unfortunately currently too demanding computationally. One week
of CPU time on a Pentium 4, 2.60GHz was needed 
to compute one density distribution in 
the $(3L\times 3L\times 3L)$-box, and one month of CPU time for the $(5L\times
5L\times 5L)$-box. Hence, a systematic study of the crossover from 3d to 2d 
in this system would require very large computational effort.  
  
\begin{figure}[t]
\begin{center}
\includegraphics[clip=,width=0.9\columnwidth]{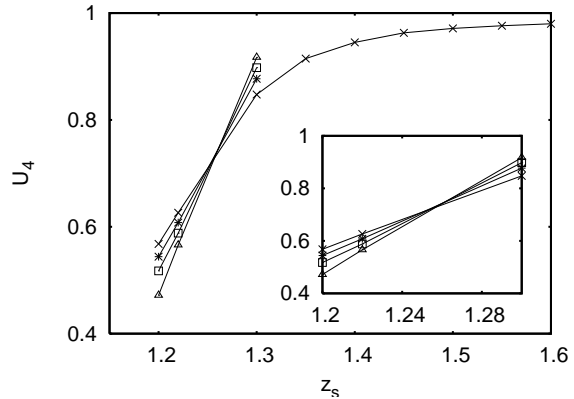} 

\caption{\label{u4ar3conf} Cumulant ratio $U_4$ as a function of the sphere
  fugacity close to the critical point for box lengths $L_x=12D$
  (crosses), $15D$ (stars), $18D$ (squares) and $21D$ (triangles) between two
  hard walls at distance $d/D = 3 L/D$. Aspect ratio $L/D=3$. The inset shows
  a magnified plot of the region near the intersection point.} 
\end{center}
\end{figure}

\subsubsection{Order parameter profiles}

The walls change the structural properties of the gas-
and liquid-like phase of rods. Here we show data
from simulations in the $NVT$-ensemble, where the volume
of the system as well as the numbers of rods and spheres are fixed. The number
of particles was chosen to match the coexistence values determined 
in the grand canonical ensemble.  Although in principle finite size effects
are different in the canonical and grand canonical ensemble
\cite{landau.binder:2000}, far enough away from the critical point this
difference can safely be neglected. 
\begin{figure}[t]
\begin{center}
\includegraphics[clip=,width=0.9\columnwidth]{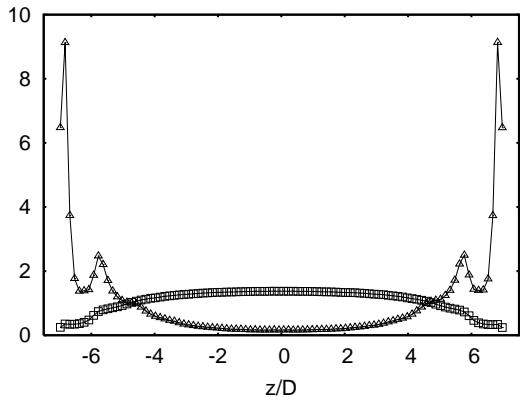} 

\caption{\label{5rhogas1.1} Density distribution of rods (triangles)
  and spheres
  (squares) normalised by their overall densities in the gas-like phase
  between two hard walls at
  distance $d/D = 3 L/D$. Aspect ratio $L/D=5$. Corresponding sphere fugacity
  is $z_s = 1.1$.} 
\end{center}
\end{figure}

\begin{figure}[t]
\begin{center}
\includegraphics[clip=,width=0.9\columnwidth]{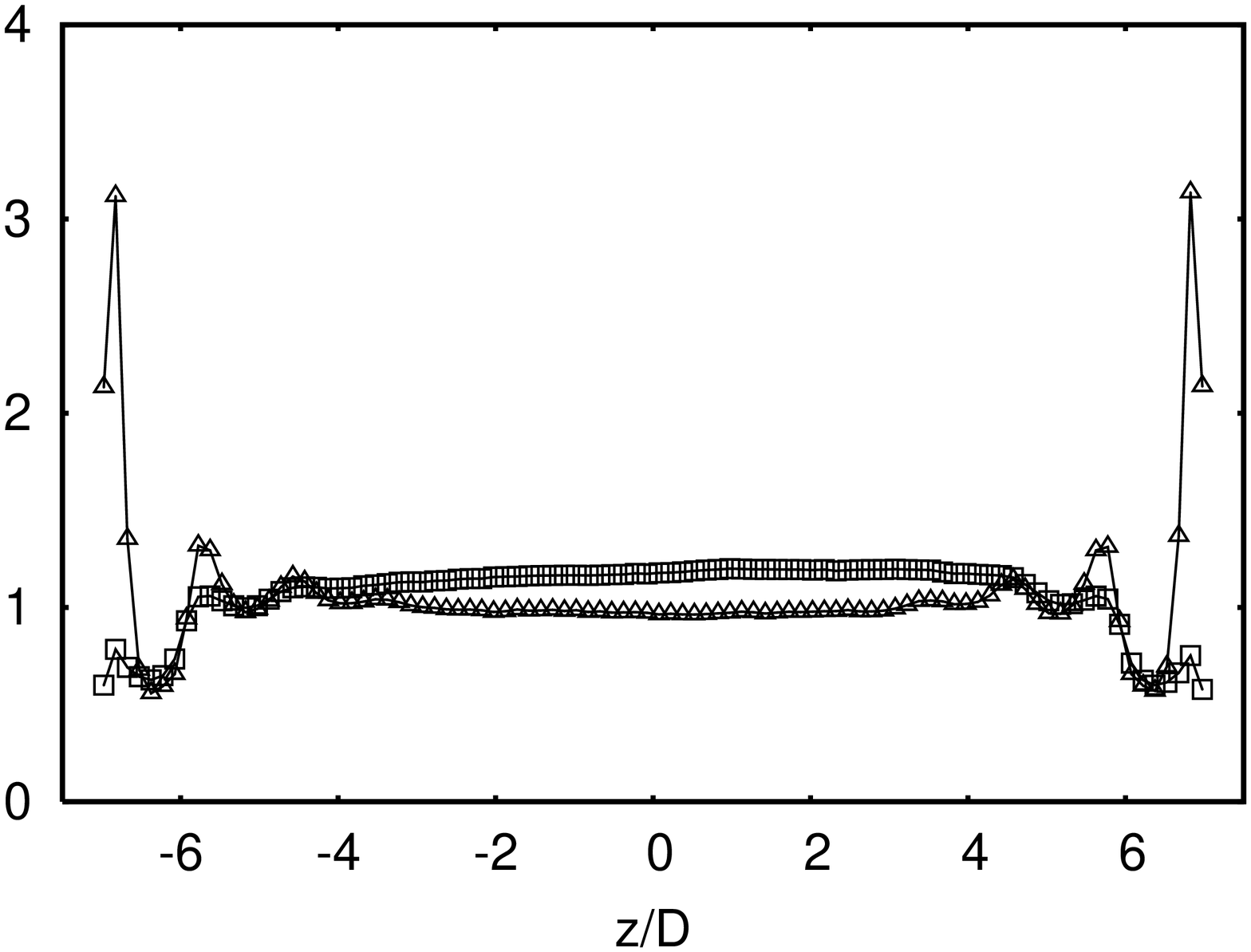} 

\caption{\label{5rholiquid1.1} Density distribution of rods
  (triangles) and spheres
  (squares) normalised by their overall densities in the liquid-like phase
  between two hard walls at
  distance $d/D = 3 L/D$. Aspect ratio $L/D=5$. Corresponding sphere fugacity
  is $z_s = 1.1$.} 
\end{center}
\end{figure}

\begin{figure}[t]
\begin{center}
\includegraphics[clip=,width=0.9\columnwidth]{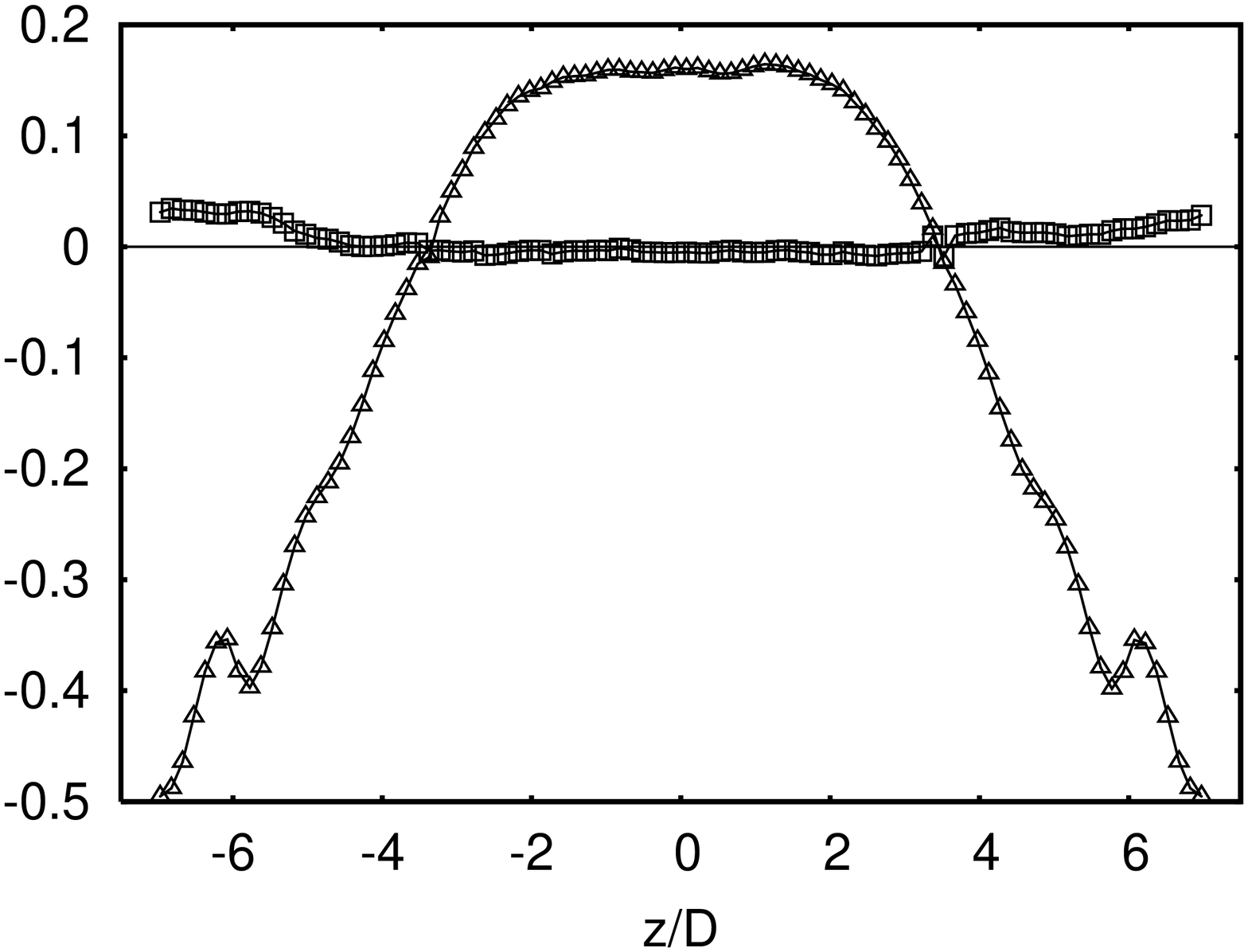} 

\caption{\label{5sbgas1.1}  Nematic order $S$ (triangles) and biaxiality $\xi$
  (squares) parameter
  distributions in the gas-like phase between two hard walls at
  distance $d/D = 3 L/D$. Aspect ratio $L/D=5$. Corresponding sphere fugacity
  is $z_s = 1.1$.} 
\end{center}
\end{figure}

\begin{figure}[t]
\begin{center}
\includegraphics[clip=,width=0.9\columnwidth]{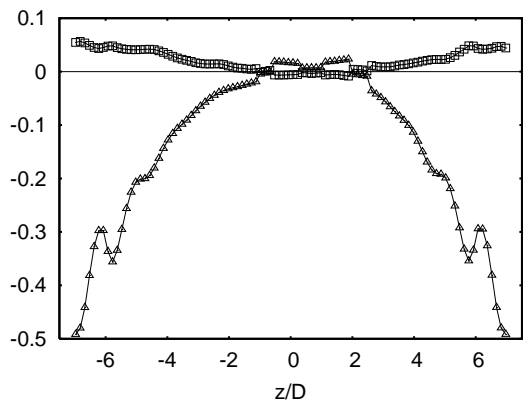} 

\caption{\label{5sbliquid1.1} Nematic order $S$ (triangles) and biaxiality $\xi$
  (squares) parameter
  distributions in the liquid-like phase between two hard walls at
  distance $d/D = 3 L/D$. Aspect ratio $L/D=5$. Corresponding sphere fugacity
  is $z_s = 1.1$.} 
\end{center}
\end{figure}

To study the anchoring effects of the walls, we define the nematic
order parameter $S$ and the
biaxiality parameter $\xi$. $S$ is the largest {\bf absolute} eigenvalue
\cite{low:2002} of the matrix 
\begin{equation}
{\bf Q} = \frac{1}{2\tilde{N}_r}\sum_{i=0}^{\tilde{N}_r} \left(3{\bf u}^i{\bf u}^i-{\bf I}\right), 
\end{equation} 
where ${\bf u}^i$ is a unit vector in the direction of the orientation of the
rod $i$ and ${\bf I}$ is the identity matrix. We divide the space between the
walls into thin slices, so $\tilde{N}_r$ is the number of rods in such a slice
at the distance $z/D$ from the middle of the simulation box. (Note that many
authors use the
largest eigenvalue instead of the eigenvalue with the largest absolute value,
which leads to different results in the case of uniaxial surface ordering!)
$S$ indicates if there is a preferred direction in the system and how 
strongly the rods are oriented with respect to it. The
eigenvector to this eigenvalue is called director. If $S$ is zero, 
the phase is completely isotropic. If $S$ is unity, all rods
are aligned parallel to the director. If $S$ is negative, they lie
perpendicular to the director. The biaxiality measure $\xi$ is half of the
difference of the other two eigenvalues of the matrix ${\bf Q}$. It shows
whether there is another preferred direction in the plane perpendicular to the
director.  

Figures \ref{5rhogas1.1} and \ref{5rholiquid1.1} show the density
distributions of rods with aspect ratio $L/D = 5$ and spheres
between walls at a
distance $d/D = 3 L/D$. Figure \ref{5rhogas1.1} shows the gas-like phase,
figure \ref{5rholiquid1.1} the
corresponding liquid-like phase at coexistence. The overall densities are
approximately the positions of the peaks of the probability distribution
$P(\eta_r)$ from the grand canonical simulation at sphere fugacity $z_s =
1.1$.

Figures \ref{5sbgas1.1} and \ref{5sbliquid1.1} show the corresponding
profiles of the nematic
order parameter $S$ and of the
biaxiality parameter $\xi$ in the gas- and liquid-like phases
respectively. 

The positional as well as the orientational order of rods are
clearly visible in the liquid- as well as in the gas-like
phase. The range of the induced effects is of the order of the rod length,
which is short in comparison to the chosen distance between the walls. In the
middle of the system the order parameters reach their bulk values. The spheres
are pushed away from the walls. This effect is also of the order of the rod
length in both phases.

\section{Discussion and summary}
\label{summary}

We have presented simulation results on the phase diagram of mixtures of hard
spherocylinders and ``penetrable hard'' spheres in the bulk and in
confinement. We hope that these results are
useful for experimental investigations with suspensions of viruses and
polymers. 

We have studied isotropic-isotropic demixing by
simulations in the grand canonical ensemble. In order to access states of high
free energy we used the successive umbrella sampling method. 
The resulting phase boundaries were compared to free volume
theory. We extracted the critical point from an analysis of the cumulants of 
the order parameter distribution. Free volume theory works well far 
away from the critical point, but, as expected, underestimates the 
concentrations at the critical point. 

In the bulk the system is very similar to the
Asakura-Oosawa-Vrij-model. In particular, we showed that its behaviour on
approach to the 
critical point falls into the Ising universality class. In confinement, 
however, the orientational degrees of freedom play a role. As the rods 
anchor parallel to the wall, the gas-like branch of the coexistence region
moves to higher colloid (rod) volume fractions than in the Asakura-Oosawa-Vrij model. And the
walls induce a much larger shift in the critical colloid (rod)
volume fraction than they do for spherical colloids.   

\acknowledgments
We would like to thank J\"urgen Horbach, Richard Vink and Peter Virnau 
for helpful 
suggestions. This work was part of the priority program SFB Tr6 
(project D5) of the German Research Association (DFG). It was partially 
funded by the DFG Emmy-Noether-Program, the
MWFZ Mainz. It has also been supported by the European Comission under the 6th
Framework Program through integrating and strengthening the European Research
Area. Contract: SoftComp VP-06/109. We thank the
Forschungszentrum J\"ulich for CPU time on the JUMP.

\end{document}